\pdfoutput=1

\documentclass{sig-alternate-2013}

\usepackage{layouts}

\usepackage{flushend}
\usepackage[numbers,sort]{natbib}
\usepackage{url}
\usepackage{tabularx}
\usepackage[iso-8859-15]{inputenx}
\usepackage{textcomp}
\usepackage{balance}
\usepackage{tabu}
\usepackage{dblfloatfix}
\usepackage[shortlabels]{enumitem}
\usepackage{verbatim}
\usepackage{amsfonts}
\usepackage{bm}
\usepackage{amssymb}
\usepackage{amsmath}
\usepackage{xspace}
\usepackage[center]{subfigure}
\usepackage{graphicx}
\usepackage{mathtools}
\usepackage{footmisc}
\usepackage{mathptmx}

\usepackage{microtype}
\usepackage{layouts}

\usepackage{mathtools}

\DeclarePairedDelimiter\floor{\lfloor}{\rfloor}

\newcommand{\footnoteurl}[1]{\footnote{\url{#1}}}

\newcommand{\approach}{\mbox{HypTrails}\xspace}

\makeatletter
\def\@copyrightspace{\@float{copyrightbox}[b]
\begin{center}
\setlength{\unitlength}{1pc}
\begin{picture}(20,3) 
\put(0,-0.95){\crnotice{\@toappear}}
\end{picture}
\end{center}
\end@float}
\makeatother

\permission{Copyright is held by the International World Wide Web Conference Committee (IW3C2). IW3C2 reserves the right to provide a hyperlink to the author's site if the Material is used in electronic media.}
\conferenceinfo{WWW 2015,}{May 18--22, 2015, Florence, Italy.} 
\copyrightetc{ACM \the\acmcopyr}
\crdata{978-1-4503-3469-3/15/05. \\
http://dx.doi.org/10.1145/2736277.2741080}

\graphicspath{{figures/}}

\begin{document}


\title{\approach: A Bayesian Approach for\\Comparing Hypotheses About Human Trails on the Web}

\numberofauthors{4}
\author{
   \alignauthor Philipp Singer\\
     \affaddr{GESIS \& Graz University of Technology}\\
     \email{philipp.singer@gesis.org}\\
   \alignauthor Denis Helic\\
     \affaddr{Graz University of Technology}\\
     \email{dhelic@tugraz.at}\\
   \and
   \alignauthor Andreas Hotho\\
     \affaddr{University of W\"urzburg}\\
     \email{hotho@informatik.uni-wuerzburg.de}\\
   \alignauthor Markus Strohmaier\\
     \affaddr{GESIS \& University of Koblenz-Landau}\\
     \email{markus.strohmaier@gesis.org}\\
}


\maketitle

\begin{abstract}

When users interact with the Web today, they leave sequential digital trails on a massive scale. Examples of such human trails include Web navigation, sequences of online restaurant reviews, or online music play lists. Understanding the factors that drive the production of these trails can be useful for e.g., improving underlying network structures, predicting user clicks or enhancing recommendations. In this work, we present a general approach called \approach for comparing a set of hypotheses about human trails on the Web, where hypotheses represent beliefs about transitions between states. Our approach utilizes Markov chain models with Bayesian inference. The main idea is to incorporate hypotheses as informative Dirichlet priors and to leverage the sensitivity of Bayes factors on the prior for comparing hypotheses with each other. For eliciting Dirichlet priors from hypotheses, we present an adaption of the so-called (trial) roulette method.
We demonstrate the general mechanics and applicability of \approach by performing experiments with (i) synthetic trails for which we control the mechanisms that have produced them and (ii) empirical trails stemming from different domains including website navigation, business reviews and online music played.
Our work expands the repertoire of methods available for studying human trails on the Web.
\end{abstract}

%
%
\vspace{1mm}
\noindent
{\bf Categories and Subject Descriptors:} H.5.3 {\textbf{[Information Interfaces and Presentation]}}: {Group and Organization Interfaces---\emph{Web-based interaction}}


\section{Introduction}
\label{sec:introduction}

The idea of human trails in information systems can be traced back to early work by Vannevar Bush ("As We May Think" \cite{bush1945we}), in which he described a hypothetical system called \emph{Memex}. Bush hypothesized that human memory operates by association, with thoughts defined by internal connections between concepts. The Memex itself was intended as users' extension of their memory, where common associative trails between documents can be stored, 
\newpage
accessed and shared. Eventually, Bush's ideas led to the concept of Hypertext \cite{nelson1965complex} and the development of the World Wide Web \cite{berners2000weaving}. 

Today, the Web facilitates the production of human trails on a massive scale; examples include consecutive clicks on hyperlinks when users navigate the Web, successive songs played in online music services or sequences of restaurant reviews when sharing experiences on the Web. Understanding such human trails and how they are produced has been an open and complex challenge for our community for years. A large body of previous work has tackled this challenge from various perspectives, including (i) modeling 
\cite{singer2014detecting,pirolli,chi2001}, (ii) regularities and patterns \cite{huberman,walk2014cikm,walk2014discovering} and (iii) cognitive strategies, finding that, for example, humans prefer to consecutively choose semantically related nodes \cite{west,singer2013computing}, humans participate in partisan sharing \cite{an} or users benefit from following search trails \cite{white2010assessing}. In this paper, we are interested in tackling an important sub-problem of this larger challenge.

\begin{figure*}
\vspace{-10pt}
\begin{center}

\subfigure[\emph{Uniform hypothesis:} all transitions are equally likely]{\label{subfig:hypo_uniform}\includegraphics[width=0.239\textwidth]{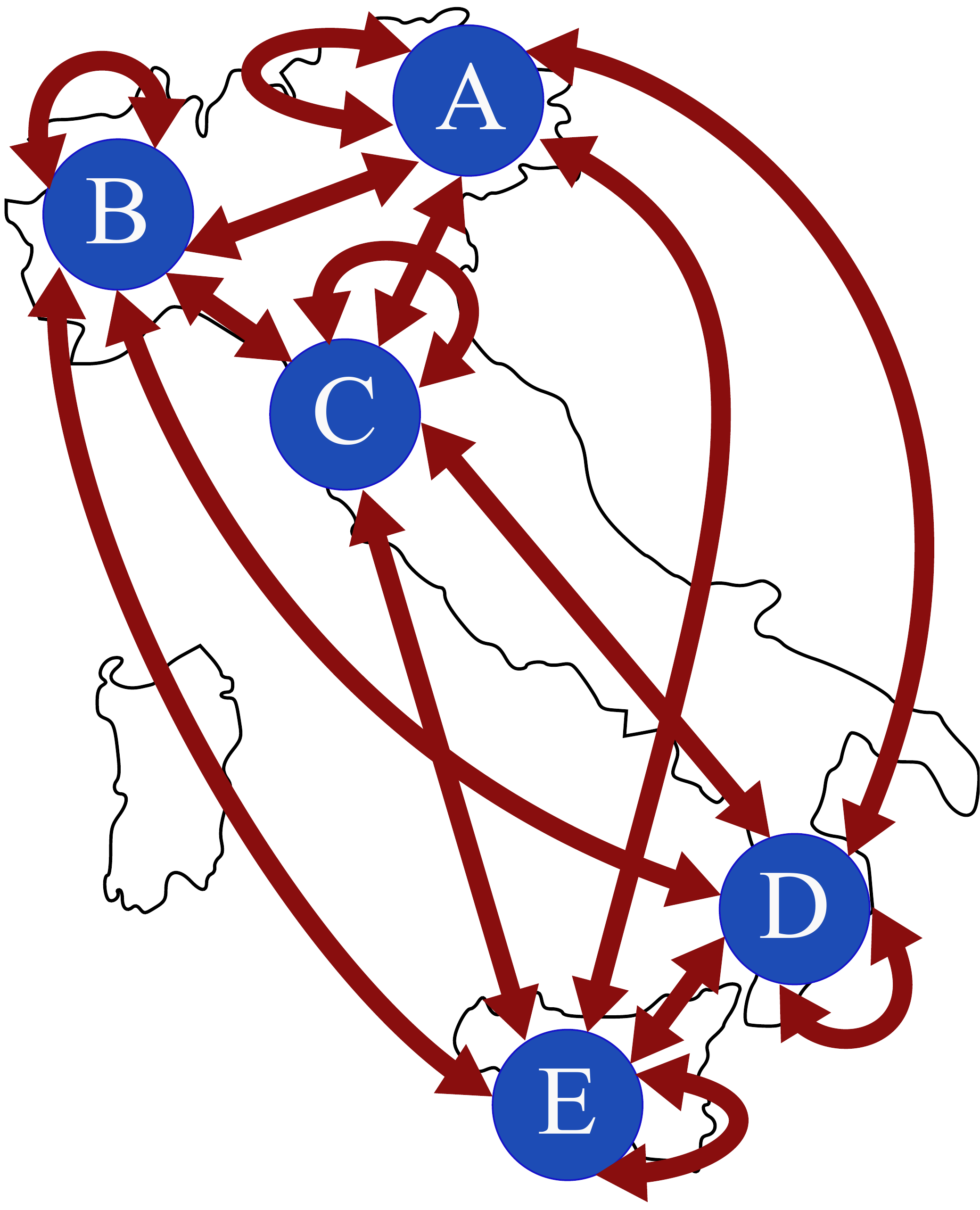}}
\hspace*{\fill}
\subfigure[\emph{Geo hypothesis:} regional node transitions are most likely]{\label{subfig:hypo_geo}\includegraphics[width=0.239\textwidth]{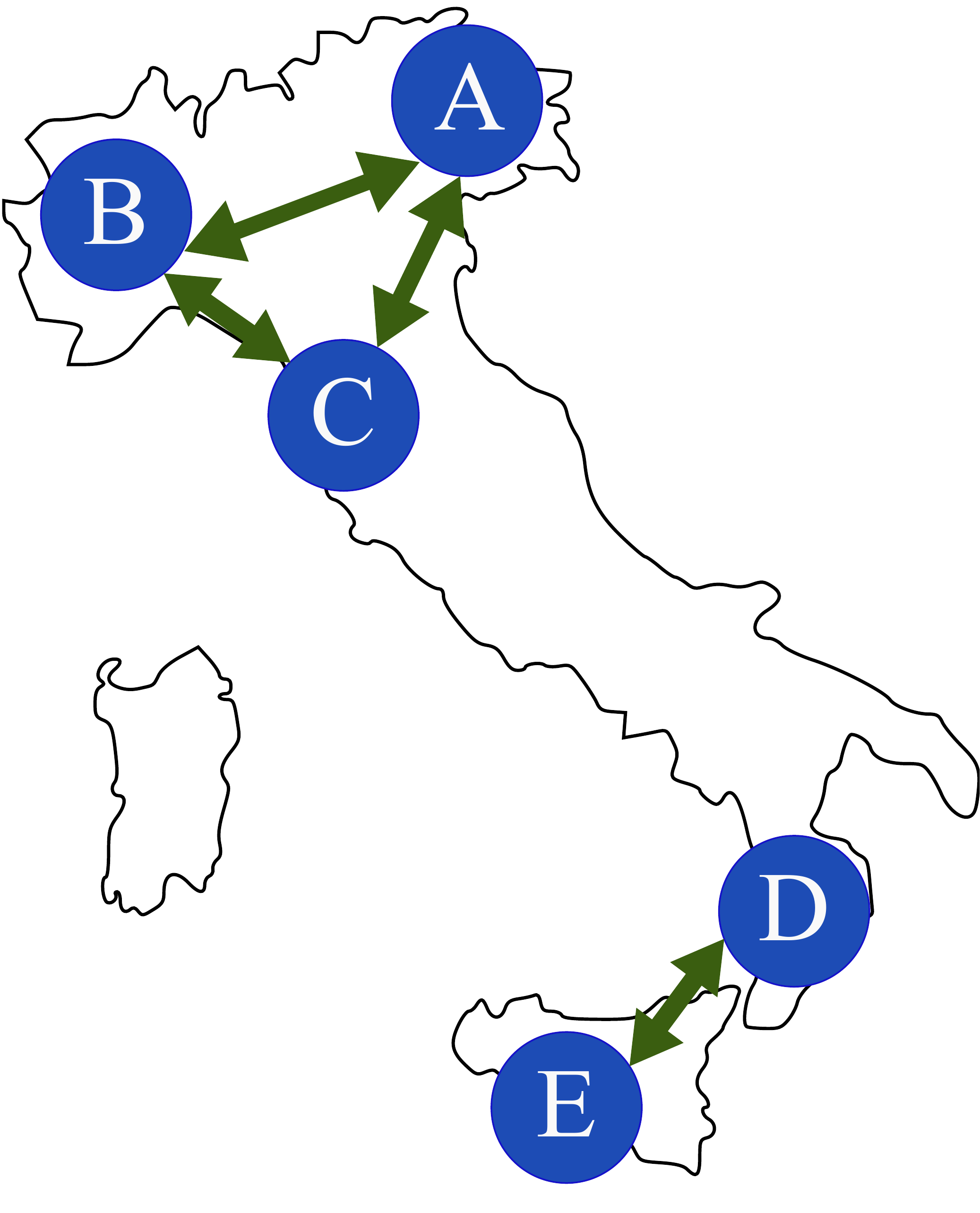}}
\hspace*{\fill}
\subfigure[\emph{Self-loop hypothesis:} self-transitions are most likely]{\label{subfig:hypo_self}\includegraphics[width=0.239\textwidth]{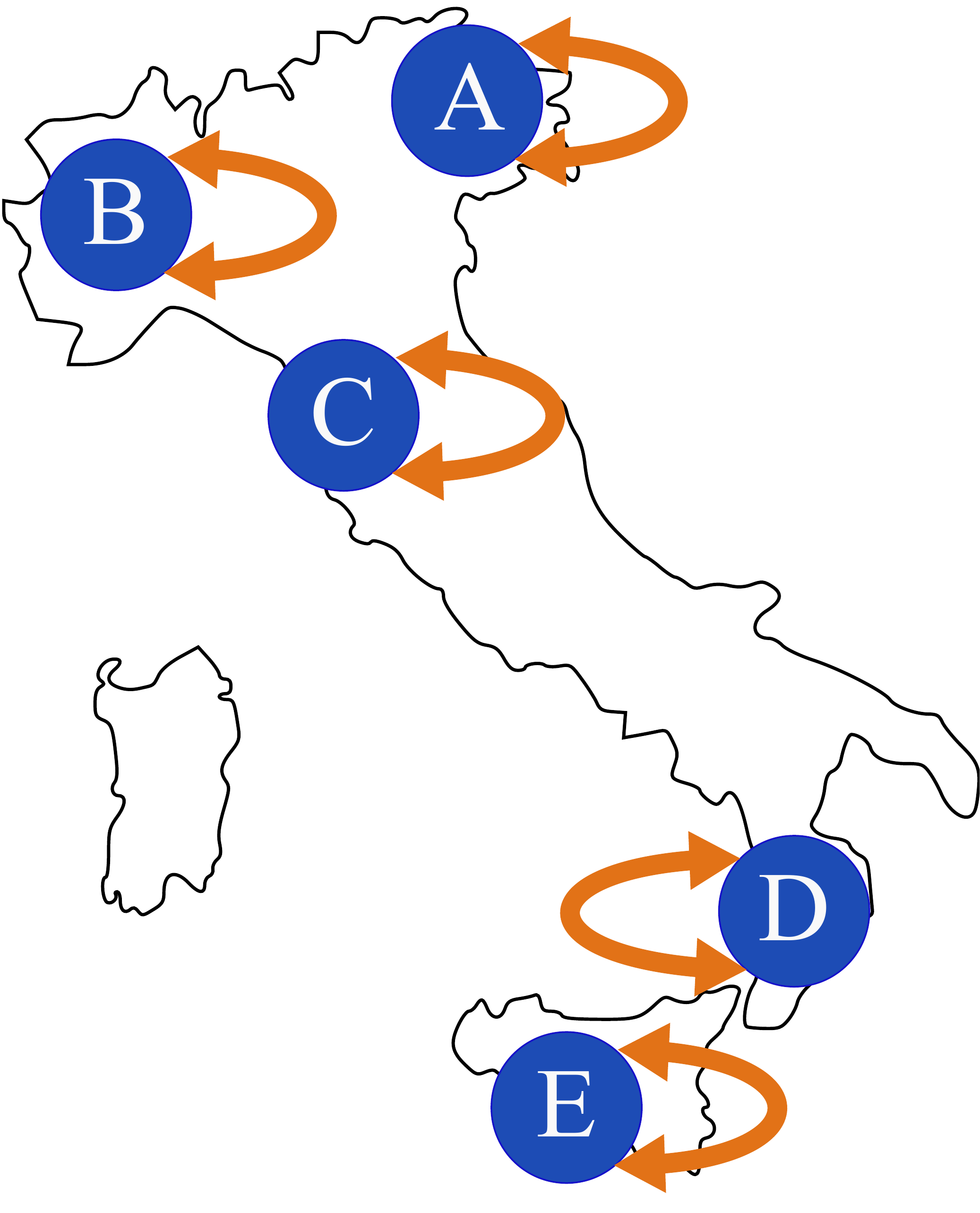}}
\hspace*{\fill}
\subfigure[\emph{Empirical transitions:} obtained from real world data]{\label{subfig:hypo_data}\includegraphics[width=0.239\textwidth]{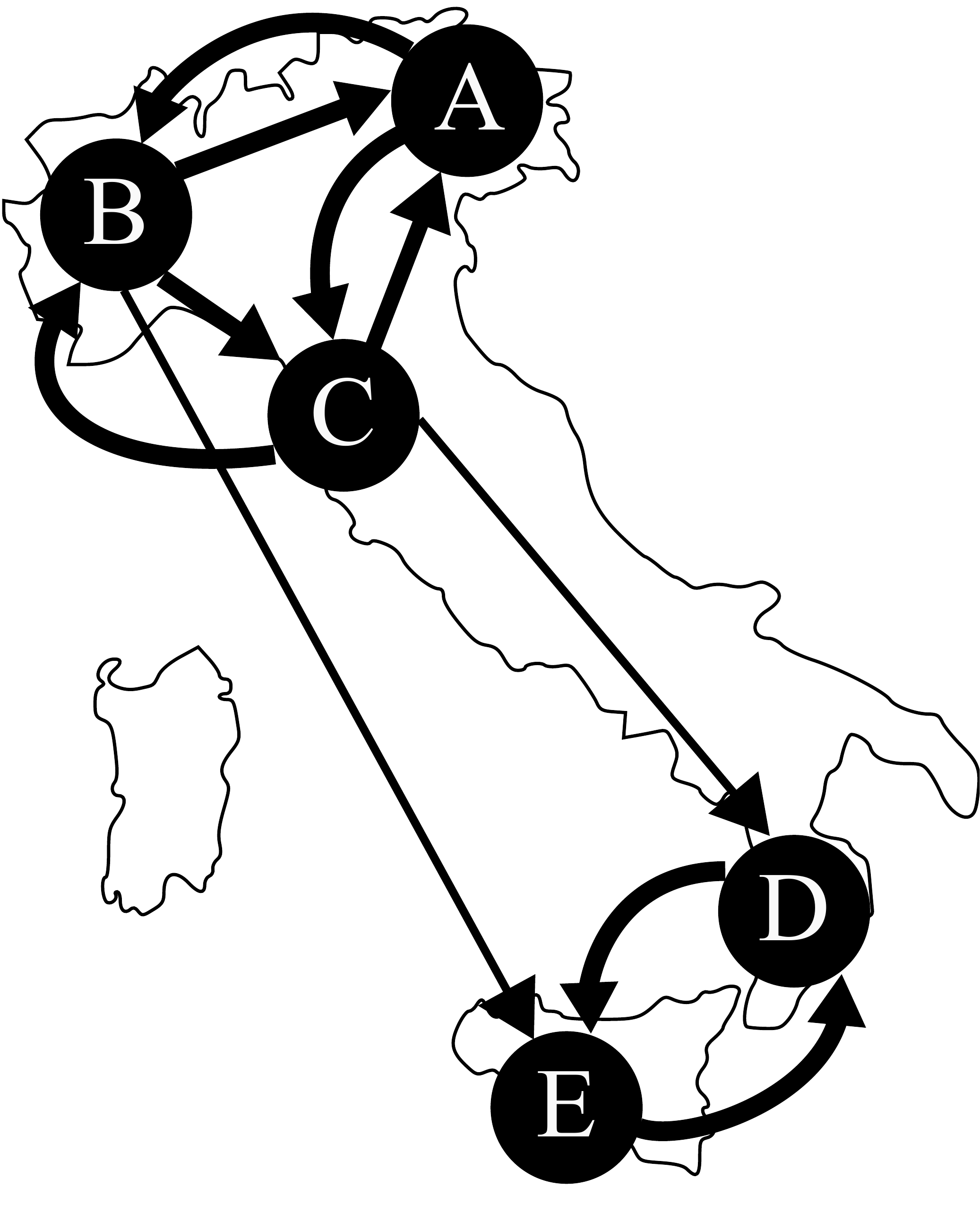}}
\end{center}
\vspace{-20pt}
\caption{\emph{Example.} This figure illustrates three exemplary hypotheses \subref{subfig:hypo_uniform}, \subref{subfig:hypo_geo}, and \subref{subfig:hypo_self} about human trails 
as well as empirical observations obtained from real-world data \subref{subfig:hypo_data}. We look at trails of online restaurant reviews in Italy; nodes A--E represent restaurants.
Hypotheses \subref{subfig:hypo_uniform}--\subref{subfig:hypo_self}
are expressed via edges, with edge weights indicating strength of belief. For empirical data \subref{subfig:hypo_data}, edge weights correspond to actually observed transitions (how many times a restaurant has been reviewed before another restaurant).
Our proposed approach compares evidences for different hypotheses given observed data \subref{subfig:hypo_data}. In this example, the geographic hypothesis \subref{subfig:hypo_geo} would be the most plausible one as we can mostly observe regional transitions between restaurants in the data \subref{subfig:hypo_data}.} 
\label{fig:hypo}
\vspace{-10pt}
\end{figure*}

\smallskip
\noindent
\textbf{Problem.}
In particular, we take a look at the problem of expressing and comparing different \emph{hypotheses} about \emph{human trails} given empirical observations. We define a trail as \emph{a sequence of at least two successive states}, and hypotheses as \emph{beliefs about transitions between states}. An intuitive way of expressing such hypotheses is in the form of Markov transitions and our beliefs in them. For example, we might have various hypotheses about how humans consecutively review restaurants on Yelp. Figure~\ref{fig:hypo} (a-c) shows three exemplary hypotheses for transitions between five restaurants (A-E) in Italy, and actual empirical transitions (d). The \emph{uniform hypothesis} in Fig \ref{fig:hypo}(a) expresses a belief that all transitions are equally likely (a complete digraph). In Fig \ref{fig:hypo}(b), the \emph{geo hypothesis} expresses a belief that humans prefer to consecutively review geographically close restaurants, while the \emph{self-loop hypothesis} in Fig \ref{fig:hypo}(c) expresses the belief that humans repeatedly review the same restaurant without ever reviewing another one. Other hypotheses are easily conceivable. What is difficult today is \emph{expressing and comparing such hypotheses within a coherent research approach}. Such an approach would allow to make relative statements about the plausibility of different hypotheses given empirical data about human trails.

\smallskip
\noindent
\textbf{Objectives.} 
We thus tackle the problem of \emph{comparing a set of hypotheses about human trails given data}. We present a Bayesian approach---which we call \emph{\approach}\footnote{Portmanteau for Hyp(ertext/otheses) Trails}---that provides a general solution to this problem. 
\newpage

\smallskip
\noindent
\textbf{Approach \& Methods.}
The \approach approach utilizes a \emph{Markov chain model} for modeling human trails and \emph{Bayesian inference} for comparing hypotheses. The main idea is to (i) let researchers express hypotheses about human trails as adjacency matrices which are then used for (ii) eliciting informative Dirichlet priors using an adapted version of the (trial) roulette method. Finally, the approach (iii) leverages the sensitivity of Bayes factors 
on the priors for comparing hypotheses with each other. 
We experimentally illustrate our approach by studying synthetic datasets with known mechanisms which we then express as hypotheses. We demonstrate the general applicability of \approach by comparing hypotheses for empirical datasets from three distinct domains (Wikigame, Yelp, Last.fm).




\smallskip
\noindent
\textbf{Contributions.}
Our main contribution is the presentation of\linebreak \approach, a general approach for expressing and comparing hypotheses about human trails. While the basic building blocks of \approach are well established (Markov chains, Bayesian inference), we combine, adapt and extend them in an innovative way that facilitates intuitive expression and elegant comparison of hypotheses. In particular, our adaption of the (trial) roulette method represents a simple way of eliciting priors for Markov chain modeling. We demonstrate the applicability of our framework in a series of experiments with synthetic and real-world data. Finally, to facilitate reproducibility and future experimentation, we make an implementation of \approach openly available to the community\footnote{\url{https://github.com/psinger/HypTrails}}.  


\smallskip
\noindent
\textbf{Structure.}
We present our approach in Section~\ref{sec:approach}. Section~\ref{sec:data} describes the synthetic and empirical data analyzed; corresponding experiments are presented in Section~\ref{sec:experiments}. We discuss our work in Section~\ref{sec:discussion}, present related work in Section~\ref{sec:relwork} and conclude in Section~\ref{sec:conclusion}.

\section{The HypTrails Approach}
\label{sec:approach}

We start with defining the problem setting and giving a short overview of the proposed approach in Section~\ref{subsec:problem}.
We proceed with explaining the fundaments of our approach based on Bayesian Markov chain modeling in Section~\ref{subsec:mc} where we also emphasize our main idea of incorporating hypotheses as Dirichlet priors and leveraging the sensitivity of Bayes factors for comparing hypotheses with each other. In Section~\ref{subsec:prior} we thoroughly discuss the process of eliciting Dirichlet priors from scientific hypotheses.


%
%

\subsection{Problem Definition \& Approach}
\label{subsec:problem}

We aim to produce a partial ordering $O$ over a set of hypotheses $\textbf{H}=\{H_1,H_2,...,H_n\}$. We base the partial order on the plausibility of hypotheses given data $D$. 
Each hypothesis $H$ describes beliefs about common transitions between nodes while data $D$ captures empirically observed human trails. A hypothesis $H$ can be expressed by an adjacency matrix $Q$ where transitions $q_{i,j}$ with strong belief receive larger values than those with lower belief.

For generating the partial ordering $O$, our \approach approach resorts to Bayesian inference utilizing a Markov chain model. We incorporate a hypothesis $H$ as informative Dirichlet priors into the inference process. For eliciting Dirichlet priors $Dir(\bm{\alpha})$ from a given hypothesis $H$ expressed as matrix $Q$---i.e., for setting corresponding hyperparameters $\alpha_{i,j}$---\approach uses an adaption of the so-called (trial) roulette method. The partial ordering $O$ is achieved by calculating marginal likelihoods $P(D|H)$ (weighted averages of likelihood, where the weights come from the parameters' prior probabilities) for competing hypotheses $H$ which we then can compare with each other by determining Bayes factors $B$.

\subsection{Bayesian Markov Chain Modeling}
\label{subsec:mc}

\approach is based on Bayesian Markov chain modeling. In the following, we only cover those fundamentals that are directly related to our approach, and point the reader to previous work \cite{singer2014detecting,Strelioff} for a more detailed treatise of the topic.

\smallskip
\noindent
\textbf{Markov chain definition.}
A Markov chain model represents a stochastic system that models transitions between states from a given state space $S=\{s_1, s_2, ..., s_m\}$ with $m=|S|$ (e.g., the distinct restaurants of our example in Figure~\ref{fig:hypo}). It amounts to a sequence of random variables $X_1, X_2, ..., X_t$. This random process is usually memoryless (the so-called Markov property, first-order) meaning that the next state only depends on the current state and not on a sequence of preceding states. Note though that Markov chain models can also be extended to incorporate higher orders; see Section~\ref{sec:discussion} for a discussion. We can define the Markov property as:

\vspace{-1em}
\begin{equation}
\begin{multlined}
 P(X_{t+1}=s_j|X_1=s_{i_1}, ..., X_{t-1}=s_{i_{t-1}}, X_t=s_{i_t})= \\ 
 P(X_{t+1}= s_j|X_t=s_{i_t}) = p_{i,j}.
\end{multlined}
\end{equation}

Markov chain models have been established as a robust method for modeling human trails on the Web in the past (e.g., \cite{singer2014detecting,chierichetti,walk2014discovering}), specifically focusing on human navigational trails (e.g., \cite{pirolli,borges1999}) with Google's PageRank being the most prominent example \cite{brin}. Hence, the Markov chain model is a natural and intuitive choice for our approach as it lets us explicitly model human trails with a dependence of the next state on the current state. We also consider hypotheses as beliefs about transitions without memory.

%

A Markov model is usually represented by a stochastic transition matrix $P$ with elements $p_{i,j}=P(s_j|s_i)$ which describe the probability of transitioning from state $s_i$ to state $s_j$; the probabilities of each row sum to $1$. The elements of this matrix are the parameters $\theta$ that we want to determine. For doing so we resort to Bayesian inference.


\smallskip
\noindent
\textbf{Bayesian inference.}
Bayesian inference refers to the Bayesian process of inferring the unknown parameters $\theta$ from data; it treats data and model parameters as random variables. For a more detailed discussion of Bayesian inference please refer to \cite{singer2014detecting,Strelioff}.  
Following Bayes' rule, the posterior distribution of parameters $\theta$ given data $D$ and hypothesis $H$ is then defined as:
\begin{equation}
 \overbrace{P(\theta| D, H)}^{\text{posterior}} = \frac{\overbrace{P(D | \theta, H)}^{\text{likelihood}}\overbrace{P(\theta|H)}^{\text{prior}}}{\underbrace{P(D|H)}_{\text{marginal likelihood}}}
 \label{eq:bayes}
\end{equation}


The \emph{likelihood function} describes the likelihood that we observe data $D$ with given parameters $\theta$ and hypothesis (model) $H$. 
The \emph{prior} reflects our belief about the parameters before we see the data or, more technically, the prior \emph{encodes} our hypothesis $H$. 
Thus, \emph{we use the prior as a representation for different hypotheses about human trails}. More precisely, we model the data with different, mostly informative, priors. We use the conjugate prior of the categorical distribution as the prior of each row of the transition matrix $P$; i.e., the Dirichlet distribution $Dir(\bm{\alpha})$. The hyperparameters $\bm{\alpha}$ represent our prior belief of the parameters and can be seen as a vector of pseudo counts $\bm{\alpha}=[\alpha_1,\alpha_2,...,\alpha_m]$. Given such a prior, the posterior distribution represents a combination of our prior belief and the actual data that we observe. For each row $i$ of $P$ we now have a posterior in the form of $Dir(n_{i,1} + \alpha_{i,1}, ..., n_{i,m} + \alpha_{i,m})$ where $n_{i,j}$ are the actual transition counts of the data between states $s_i$ and $s_j$ and $\alpha_{i,j}$ are the prior pseudo counts assigned to this transition. We provide a thorough description of how we elicit the needed Dirichlet priors from expressed hypotheses by researchers in Section~\ref{subsec:prior}. 


\smallskip
\noindent
\textbf{Comparing hypotheses.}
Finally, the \emph{marginal likelihood} (which we can also call evidence) expresses the probability of the data given a hypothesis $H$ and plays a crucial role for comparing hypotheses;\footnote{\label{fn:evidence}We calculate log-evidence utilizing logarithms of the gamma function for avoiding underflow.} it is defined as follows (for derivation please consult \cite{singer2014detecting,Strelioff}):

\begin{equation}
 P(D | H) = \prod_i\frac{\Gamma(\sum_j \alpha_{i,j})}{\prod_j \Gamma(\alpha_{i,j})} \frac{\prod_j \Gamma(n_{i,j}+\alpha_{i,j})}{\Gamma(\sum_j (n_{i,j}+\alpha_{i,j}))}
\label{eq:evidence}
\end{equation}
Note that the hyperparameters $\alpha_{i,j}$ differ for various hypotheses $H$ as we express them via different Dirichlet priors; the actual transition counts $n_{i,j}$ are the same for each hypothesis. 
For comparing the plausibility of two hypotheses, we resort to \emph{Bayes factors} \cite{kass1995bayes,wasserman2000bayesian}. 
Bayes factors are representing a Bayesian method for model comparison that include a natural \emph{Occam's razor} guarding against overfitting. In our case, a model represents a hypothesis at interest with each having different priors with different hyperparameters that express corresponding beliefs. For illustrative purposes, we are now interested in comparing hypotheses $H_1$ and $H_2$ where $H_1,H_2\in\textbf{H}$, given observed data $D$. We can define the Bayes factor---note that we apply unbiased comparison assuming that all hypotheses are equally likely a priori---as follows:

%



\begin{equation}
B_{1,2} = \frac{P(D | H_1)}{P(D|H_2)}
\end{equation}


$P(D|H)$ is the marginal likelihood (evidence) defined in Equation~\ref{eq:evidence} and the Bayes factor
can be seen as a summary of the evidence provided by the data in favor of one scientific hypothesis over the other.
\approach is not only suited for comparing two hypotheses with each other, but rather a set of hypotheses $\textbf{H}=\{H_1,H_2,...,H_n\}$.
For determining the partial order $O$ over $\textbf{H}$, we order the evidences that data $D$ provides in favor of hypotheses $H$; i.e., by ordering $P(D|H)$ using a less-than-equal binary relation. However, ordering the evidences is not enough as we need to check the significance of their ratios which we tackle by calculating Bayes factors. In case that the significance is not present, we consider two hypotheses as being equal.
 For determining the strength of the Bayes factor we resort to Kass and Raftery's \cite{kass1995bayes} interpretation table. 



\smallskip
\noindent
\textbf{Leveraging the sensitivity of Bayes factors.} 
Throughout this section we have described our main idea of incorporating hypotheses in the form of informative Dirichlet priors into the inference process. We leverage marginal likelihoods and Bayes factors for making informed decisions about the relative plausibility of given hypotheses.
Usually, one common critique of Bayes factors is that they are highly sensitive with regard to the choice of the prior \cite{kass1995bayes}. In contrast, posterior measures ignore the influence of the prior the more data one observes which is why they are more ignorant to the choice of prior \cite{vanpaemel2010prior}.
In our approach, we exploit the sensitivity of Bayes factors on the prior as \emph{an elegant solution to the problem of comparing hypotheses}.  As we express our different hypotheses in the form of priors, we are explicitly interested in using a measure that is sensitive to the choice of priors and hence, can give us insights into the relative plausibility of each hypothesis. Thus, in this case, marginal likelihoods and Bayes factors are an appropriate measure for comparing scientific hypotheses \cite{vanpaemel2010prior}.

\begin{figure*}[t!]
\vspace{-10pt}
\centering
\subfigure[Uniform prior]{\label{subfig:toy_alpha}\includegraphics[width=0.33\textwidth]{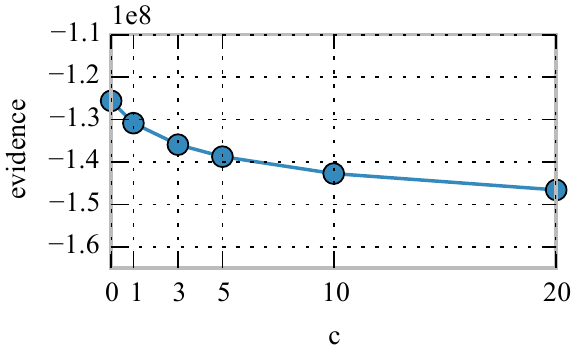}}
\subfigure[Empirically aligned prior]{\label{subfig:toy_empirical}\includegraphics[width=0.33\textwidth]{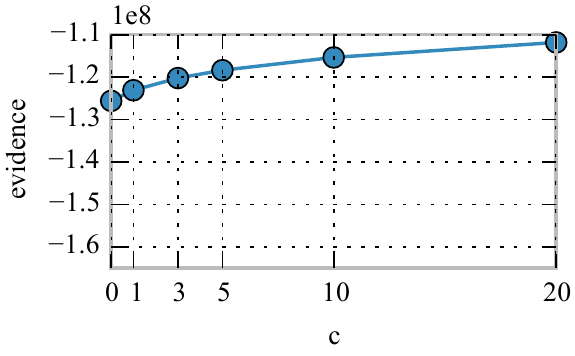}}
\subfigure[Empirically opposing prior]{\label{subfig:toy_max}\includegraphics[width=0.33\textwidth]{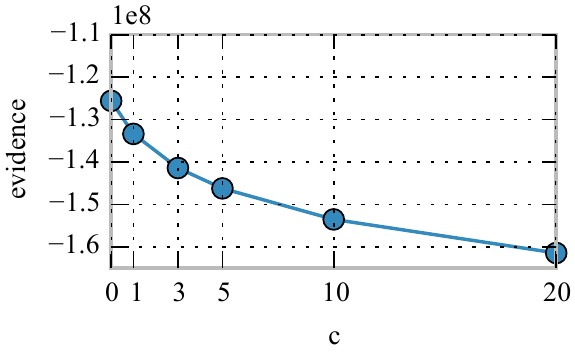}}
\vspace{-10pt}
\caption{\emph{Understanding influence of priors.} This figure shows how the choice of prior pseudo counts influences evidence; we apply several toy priors to navigation data (Wikigame, see Section \ref{sec:data}). In \subref{subfig:toy_alpha} we use a uniform Dirichlet prior which means that for each row $i$, $\bm{\alpha}$ has the same value for each element: $\alpha_{i,j}=1+c, \forall i,j$.  By increasing the constant $c$ (x-axis), we can observe that the evidence (y-axis) is decreasing; the largest evidence is at $c=0$. By using an empirically aligned prior \subref{subfig:toy_empirical} as $(n_{i,j}>0 \to\alpha_{i,j}=1+c) \wedge (n_{i,j}=0 \to \alpha_{i,j}=1), \forall i,j $, we end up with a larger evidence the larger $c$ is. Finally, in \subref{subfig:toy_max} we intentionally set "bad" prior counts for the $\bm{\alpha}$ values via $(n_{i,j}=0 \to \alpha_{i,j}=1+c) \wedge (n_{i,j}>0 \to \alpha_{i,j}=1), \forall i,j$ showing that the evidence becomes smaller as we increase $c$. The results indicate that the more a hypothesis is aligned with empirical data, the larger the evidence is, and vice versa.
}
\label{fig:toy}
\vspace{-10pt}
\end{figure*}


\subsection{Eliciting Dirichlet Priors}
\label{subsec:prior}

This section explains in greater detail how we can express the hypotheses about human trails and how \approach elicits proper informative Dirichlet priors from these.
First, we illustrate how the prior influences the evidence by studying several toy examples. Next, we present an intuitive way of eliciting the Dirichlet priors by introducing an adaption of the so-called \emph{(trial) roulette method}. 



\smallskip
\noindent
\textbf{Understanding influence of priors.}
In Section~\ref{subsec:mc}, we discussed that we use the sensitivity of Bayes factors with regard to the choice of prior (i.e., determined via marginal likelihoods) as a feature (or solution) rather than a limitation. It allows us to model hypotheses in the form of prior distributions which then can be compared by corresponding Bayes factors. But how exactly does the prior influence evidence (marginal likelihood)? Note that the posterior probability (see Equation~\ref{eq:bayes}) is a combination of our prior belief (pseudo counts) and the data we observe (transition counts) which is why both influence the evidence (see Equation~\ref{eq:evidence}).


To illustrate the influence, we apply several toy priors to human trail data. The choice of data is secondary and we observe the same behavior regardless of the underlying data; in this case we exemplary use 
navigation data (Wikigame dataset introduced in Section~\ref{sec:data}). First, we apply a \emph{uniform prior}; i.e., $\bm\alpha$ has the same value for each row $i$ and element $j$: $\alpha_{i,j}=1+c, \forall i,j$.
By ranging the constant $c$ over $0,1,3,5,10,20$, we observe decreasing evidence (Figure~\ref{subfig:toy_alpha}) which is not surprising as the uniform pseudo counts do not mirror the observed transition counts well. Technically, with increased $c$ the Dirichlet prior concentrates more and more of its probability mass on a uniform distribution of parameters, and thus, the weights of alternative parameter configurations become smaller. However, the likelihood is larger for the alternative parameter configurations (coming from the data) and this results in a smaller weighted average of the likelihood, i.e., in a smaller evidence.

On the other hand, if we provide some form of an \emph{empirically aligned prior} as $(n_{i,j}>0 \to\alpha_{i,j}=1+c) \wedge (n_{i,j}=0 \to \alpha_{i,j}=1), \forall i,j $ 
we end up with a larger evidence the larger $c$ is as we can see in Figure~\ref{subfig:toy_empirical}. This is because we actually increase the pseudo counts of transitions that we also observe in our data while we keep the pseudo counts for non-observed transitions at $1$. In this case, we concentrate the prior probability mass on the parameter configuration that is very well aligned with the observations. As a consequence we give more weight for parameter configurations where the likelihood is anyhow large and this increases the evidence.

Finally, we illustrate the behavior of a toy prior that expressed an \emph{empirically opposing prior} in the form of $(n_{i,j}=0 \to \alpha_{i,j}=1+c) \wedge (n_{i,j}>0 \to \alpha_{i,j}=1), \forall i,j$. In this example we intentionally set the prior pseudo counts to the opposite of what the actual data tells us. 
We assign low prior pseudo counts ($1$) to elements with large observed transition counts while we incrementally increase the pseudo counts of transitions that we do not observe in our data. As expected, Figure~\ref{subfig:toy_max} shows that the evidence decays as we increase $c$. Technically, we assign the greatest weights for parameter configurations with the smallest likelihoods resulting in a steep decay of evidence with increasing values of $c$.

Note that $c=0$ 
results in the same evidence for all three toy priors as in all cases $\alpha_{i,j}=1, \forall i,j$. This refers to a Dirichlet prior that is uniformly distributed over the simplex which is a special form of a uniform prior and which we will refer to as a \emph{flat prior}.

These toy examples show that if the prior is well aligned with data, the evidence is rising with the strength of the prior. 
The evidence is the largest if the prior and the likelihood are concentrated over the same parameter regions and it is the lowest if they concentrate on different regions \cite{xie2010improving}. Hence, we want to choose an informative prior that captures the same regions as the likelihood. This leads to the observation that 
if the prior choice represents a valid hypothesis about behavior producing human trails on the Web, the evidence should be larger than a uniform prior, or an unlikely hypothesis prior with an equal amount of pseudo counts. Thus, for fair comparison, we want to compare hypotheses with each other that exhibit the same amount of pseudo counts assigned.
Next, we elaborate the (trial) roulette method for eliciting Dirichlet priors from  hypotheses.



\begin{figure*}[t!]
\vspace{-10pt}
\centering
{\includegraphics[width=0.9\textwidth]{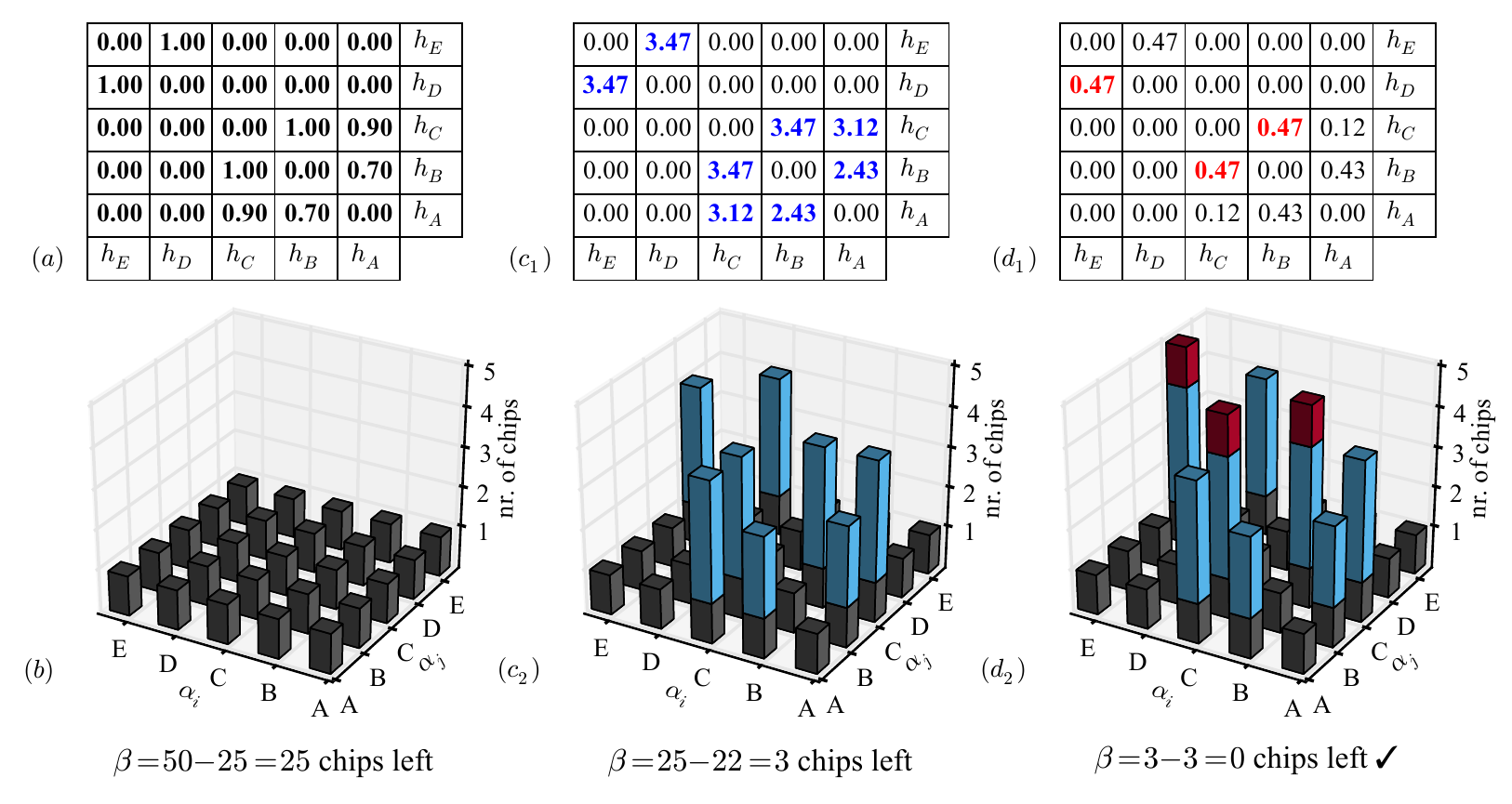}}
\vspace{-10pt}
\caption{\emph{Illustration of the (trial) roulette method.} In this figure the most important steps of our trial roulette method are visualized. We begin with a matrix expressing a researcher's hypothesis about human trails in (a)---in 
this case the exemplary geographic hypothesis for trails over businesses reviewed (cf. Figure~\ref{fig:hypo}).
The (trial) roulette method proceeds with distributing a given number of chips (pseudo counts, in this case $\beta=50$) to the Dirichlet priors. It starts by assigning one chip to each element (uniform) as can be seen in (b) before it proceeds by assigning the remaining chips according to their values of our given matrix in (a) as can be seen from (c) to (d). 
In each column, values of the matrices that receive at least one chip are marked bold and in the same color as the bars indicating chip assignments for the Dirichlet priors.
In case of ties the ranking is produced in random fashion. For details please refer to the elaboration in Section~\ref{subsec:prior}.}
\label{fig:prior_viz}
\vspace{-10pt}
\end{figure*}


\smallskip
\noindent
\textbf{(Trial) roulette method.}
Our approach requires to define the hyperparameters of prior Dirichlet distributions by setting the pseudo counts  $\alpha_{i,j}$ given the hypothesis at interest. However,
the process of eliciting prior knowledge is no trivial problem and requires careful steps (see \cite{oakley2010eliciting,garthwaite2005statistical} for a discussion). 
As a solution, we present an adaption of the so-called \emph{(trial) roulette method} which was originally proposed in \cite{gore1987biostatistics} and further discussed in \cite{oakley2010eliciting,Davidson2013}. It is a graphical method that allows experts to express their subjective belief by distributing a fixed set of chips (think about casino chips you set on a roulette table) to a given grid (e.g., bins representing result intervals). The number of chips assigned to an element of the grid then reflect the experts' belief in the specific bin. 

In our work we adapt the (trial) roulette method. The grid can be understood as a matrix $Q$ where each element $q_{i,j}$ of the grid represents the belief of a given hypothesis about the transition from state $s_i$ to state $s_j$. 
Values $q_{i,j}$ are set by researchers for expressing a hypothesis. They need to be positive values and larger values indicate stronger belief in a given transition.
The prior of each row of the transition matrix $P$ of the Markov chain model is defined as a Dirichlet distribution (cf. Section~\ref{subsec:mc}) with hyperparameters (pseudo counts) $[\alpha_{i,1}, \alpha_{i,2}, ..., \alpha_{i,j}]$  which we want to set given the hypothesis. Concretely, we want to automatically distribute a number of chips to the given pseudo counts (another grid) according to the values provided in matrix $Q$ expressing a hypothesis $H$. We define the overall number of chips to distribute for a given hypothesis as:

\vspace{-1.5em}
\begin{equation}
\beta = \overbrace{m^2}^{\text{flat prior}} + \underbrace{k\cdot m^2}_{\text{additional informative prior}}
\label{eq:chips}
\end{equation}

\vspace{-0.5em}
$m=|S|$ and $m^2$ amounts to the flat prior---i.e., we assign the same number of pseudo counts ($1$) to each transition---which is why the number of uniformly assigned chips is equal to the overall number of parameters of the Markov chain model. Additionally, we distribute $k*m^2$ informative pseudo clicks for the given hypothesis, where $k$ describes a weighting factor for the informative part. The larger we set $k$, the more we concentrate the Dirichlet distributions according to a hypothesis at interest---see Section~\ref{sec:discussion} for a discussion.

By and large, the goal of our adaption of the (trial) roulette method is to not only give researchers an intuitive way of expressing their hypotheses as matrices $Q$, but also to elicit informative Dirichlet distributions according to the values $q_{i,j}$ of $Q$.
Next, we want to illustrate the process of expressing a hypothesis and assigning prior pseudo counts via the example shown in Figure~\ref{fig:prior_viz} using the (trial) roulette method. 
Let us again focus on human trails over reviewed restaurants in Italy (see Figure~\ref{fig:hypo}) and illustrate the method using the geographic hypothesis given in Figure~\ref{subfig:hypo_geo}.
For this visualization, we assume that we set $k=1$ leading to $\beta=m^2 + m^2$ chips we want to distribute. The following steps are necessary:

\emph{\textbf{(a) Expressing the hypothesis.}} Researchers start with expressing the hypothesis matrix $Q$ with elements $q_{i,j}$ that capture the belief about transitions of underlying human trails; the matrix can be seen in Figure~\ref{fig:prior_viz}(a). 
In this example, we have five states (restaurants)---i.e., $S={A,B,C,D,E}$ and $m = 5$ leading to $\beta=50$ chips to distribute---and we express our geographic hypothesis about common transitions with values between $0$ and $1$. The precondition is that only positive values are used and larger values ($q_{i,j}$) always express stronger beliefs compared to smaller values. In this case, the closer a value is to $1$, the closer two restaurants are geographically and the more we believe in corresponding transitions. 
As can be seen in Figure~\ref{subfig:hypo_geo}, both restaurant pairs B-C and E-D are the closest in geographical terms which is why we lay the strongest belief in these symmetric transitions; concretely, we set $q_{B,C}=1.0$, $q_{C,B}=1.0$, $q_{D,E}=1.0$ and $q_{E,D}=1.0$. The next closest restaurant pair is A-C which is why we also have strong beliefs in humans consecutively reviewing restaurant A after C and vice versa; we set $q_{A,C}=0.9$ and $q_{C,A}=0.9$. Finally, we set $q_{A,B}=0.7$ and $q_{B,A}=0.7$ as we also have some (lower) belief that humans consecutively review given restaurants. We set all other transitions to zero as we believe that given restaurants are too far away.
%
%
One matrix $Q$ represents one general hypothesis $H$ about human trails---in a more realistic scenario one would want to express several of such matrices (such as for the other hypotheses given in Figure~\ref{fig:hypo}). Also note that we hand-pick values for this example and in a more rigorous investigation one would potentially use an automatic method for determining them (as we do in Section~\ref{sec:experiments}). The next steps are automatically performed by \approach for eliciting Dirichlet priors from such matrices, more specifically by the adapted (trial) roulette method.

\emph{\textbf{(b) Initial uniform distribution.}} The (trial) roulette method starts with assigning uniform chips to each transition which can be  seen as obtaining a flat prior ($\alpha_{i,j} = 1, \forall i,j$) and accounts to the left part ($m^2$) of Equation~\ref{eq:chips}. The updated prior (i.e., hyperparameters for the Dirichlet distributions) can be seen in Figure~\ref{fig:prior_viz}(b) with black bars; all elements where one chip is assigned to are marked bold and black in Figure~\ref{fig:prior_viz}(a). By subtracting the distributed number of chips from $\beta$ we have $\beta=50-25=25$ chips left for the informative part described next.

\emph{\textbf{(c) Informative distribution. }} Matrix $Q$ gets normalized and then multiplied by the number of chips left: $Q = \frac{Q}{||Q||_1} * \beta$ where $||Q||_1$ is the $\mathcal{\ell}_1$-norm and $\beta=25$. The resulting matrix can be seen in Figure~\ref{fig:prior_viz}(c$_1$). The method assigns as many chips to elements of the prior as the integer floored values of $Q$ specify. So e.g., $q_{A,B}=2.43$ and $\floor*{q_{A,B}}=2$ leading to $\alpha_{A,B}+=2$ whereas $\floor*{q_{B,D}}=0$ which is why the pseudo count for this transition is not increased. Overall, the method distributes $22$ more chips marked bold and blue in Figure~\ref{fig:prior_viz}(c$_1$) leading to $\beta=25-22=3$ chips left; the updated prior distributions (new chips marked blue) can be seen in Figure~\ref{fig:prior_viz}(c$_2$).

\emph{\textbf{(d) Remaining informative distribution.}} Finally, the method subtracts the integer floored values from $Q$ leading to the matrix illustrated in Figure~\ref{fig:prior_viz}(d$_1$) calculated by $Q = Q - \floor*{Q}$. It now needs to distribute the chips left (three in this case) according to the remaining values in $Q$. The method accomplishes this by ranking the values in descending order and assigning one chip to each element until none is left, starting from the largest and ending at the smallest. In case of a tie the ranking for the ties is produced in random fashion; hence, in this case $\alpha_{D,E}$ does not receive one more chip. We mark the elements that receive one further chip bold and red in Figure~\ref{fig:prior_viz}(d$_1$) and update our prior pseudo counts as can be seen in Figure~\ref{fig:prior_viz}(d$_2$) also in red color. Now, the (trial) roulette method has no chips left and is finished. 

The final chip assignment shown in Figure~\ref{fig:prior_viz}(d$_2$) represents the prior (hypothesis). In detail, each row corresponds to a Dirichlet distribution with corresponding pseudo counts (hyperparameters) $\alpha_{i,j}$ (e.g., $\alpha_{C,B}=5$). By proceeding, our \approach approach now uses these Dirichlet priors for Bayesian Markov chain modeling inference as described in Section~\ref{subsec:mc}. Concretely, in combination with the transitions $n_{i,j}$ observed from data they influence the marginal likelihood calculated as defined in Equation~\ref{eq:evidence}. See Figure~\ref{fig:toy} for a visualization of how the prior influences the evidence. By repeating the trial roulette method and evidence calculation for each hypothesis at interest, we can determine corresponding evidences and Bayes factors for comparing hypotheses with each other.

\section{Description of datasets}
\label{sec:data}

In this section, we introduce both synthetic as well as empirical datasets which we consider for our experiments. We produce \emph{synthetic data} consisting of simulated human trails---in this case navigational trails---with known mechanisms from a generated (i.e., artificial) network. The introduced \emph{empirical data} stems from three real-world datasets from different domains. The state space $S$ investigated is always defined by the distinct elements the trails traverse over; e.g., if we observe trails over five distinct restaurants being reviewed (see Figure~\ref{fig:hypo}) we consider these five for the state space. As described in \cite{singer2014detecting}, we additionally add a reset state to the trails and state space of each dataset to ensure ergodicity.


\subsection{Synthetic Datasets}
\label{subsec:synthetic}

We start by generating a directed random network using a generalized version of \emph{Price's preferential attachment scale-free network model} \cite{price1976general,barabasi1999emergence}. The network generation algorithm starts with a clique containing $11$ nodes and proceeds to add nodes with an out-degree of $10$ leading to an overall network size of $10,000$ nodes. These parameters are arbitrary and could be set differently. Next, we simulate three different kinds of navigational trails, each consisting of exemplary $1,000$ trails of length $5$, as follows:

\smallskip
\noindent
\textbf{Structural random walk.} 
For each trail we start at a random node of the network and perform a random walk through the network. The walker chooses the next node by randomly selecting one out-going link of the current node.

\smallskip
\noindent
\textbf{Popularity random walk.}
Again, the walker starts at a random node of the network, but now selects the next node by choosing the out-link according to the target's in-degree. The walker lays a softmax-like smoothing over the in-degrees of all target nodes ($e^{\text{deg}^-(s)/10}$); it then chooses the next node according to given probability leading to a small stochastic effect. This is aimed at averting too long loops that would happen with simple greedy selection.

\smallskip
\noindent
\textbf{Random teleportation.}
Again, we start with a random node in the network for each trail. However, we now completely ignore the underlying topological link network and simply randomly choose any other node of the network---i.e., teleporting through the network. 

\subsection{Empirical Datasets}
\label{subsec:empirical_data}

For our experiments we also consider three different empirical datasets which are described next.

\smallskip
\noindent
\textbf{Wikigame dataset.} 
First, we study navigational trails over Wikipedia pages that are consecutively visited by humans. 
The dataset is based on the online game called Wikigame (\url{thewikigame.com}) where players aim to navigate to a given Wikipedia target page starting from a given Wikipedia start page using Wikipedia's link structure only. All start-target pairs are guaranteed to be connected in Wikipedia's topological link network and users are only allowed to click hyperlinks and use the browser's buttons such as refresh, but not use other features such as the search field. In this article, we study trails collected from users playing the game between 2009-02-17 and 2011-09-12. Overall, the dataset consists of $1,799,015$ trails---where each trail represents the consecutive websites visited by one user for one game played---through Wikipedia's main namespace including $360,417$ distinct pages with an average trail length of around $6$.
We use corresponding textual and structural Wikipedia article data for hypotheses generation. In particular, we use the Wikipedia dump dated on 2011-10-07\footnote{This Wikipedia dump closely resembles the information available to players of the game for our given time period.}.

\smallskip
\noindent
\textbf{Yelp dataset.}
Second, we study human trails over successive businesses reviewed by users on the reviewing platform \emph{Yelp} (\url{yelp.com})---we have used this setting as an example throughout this article (e.g., see Figure~\ref{fig:hypo}). For generating these trails we use a dataset publicly offered by Yelp\footnote{\url{yelp.com/dataset_challenge}}. Overall, we generate $125,365$ trails---where each trail describes the subsequent review history of one single user---over $41,707$ distinct businesses with an average trail length of $8$.
The data also includes further information about the businesses like the geographic location or category markers assigned, which we will use for hypotheses generation.

\smallskip
\noindent
\textbf{Last.fm dataset.} 
Third, we study human trails that capture consecutive songs listened to by users on the music streaming and recommendation website
\emph{Last.fm} (\url{lastfm.com}).  We use a publicly available dataset \cite{celma} for generating the trails at hand focusing on data from one day (2009-01-01). Overall, the dataset consists of $275$ trails---where each trail captures the successive songs listened to by one user on a given day---over $11,166$ distinct tracks with an average trail length of $52.8$. For generating hypotheses, we consult the \emph{MusicBrainz} (\url{musicbrainz.org}) API as described later.

\begin{figure*}[t!]
\vspace{-10pt}
\centering
\subfigure[Structural random walk]{\label{subfig:synthetic_uniform}\includegraphics[width=0.33\textwidth]{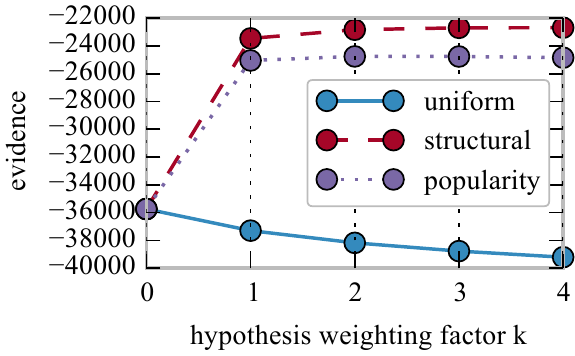}}
\subfigure[Popularity random walk]{\label{subfig:synthetic_popularity}\includegraphics[width=0.33\textwidth]{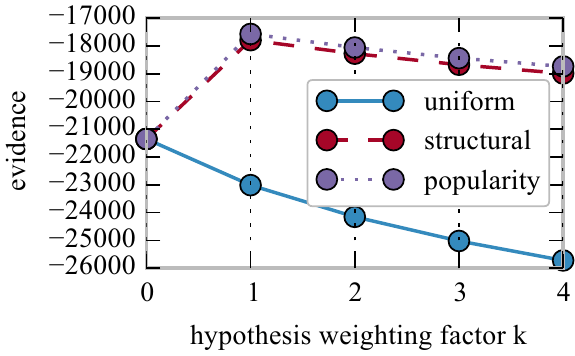}}
\subfigure[Random teleportation]{\label{subfig:synthetic_teleport}\includegraphics[width=0.33\textwidth]{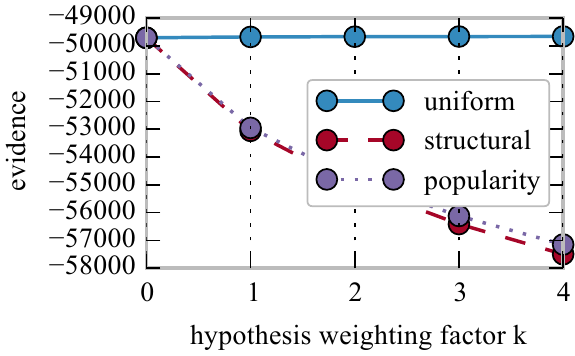}}
\vspace{-10pt}
\caption{\emph{Experiments with synthetic data.} This figure depicts the results obtained when applying \approach to three synthetically generated trail corpora with known mechanisms (structural random walk \subref{subfig:synthetic_uniform}, popularity random walk \subref{subfig:synthetic_popularity} and random teleportation \subref{subfig:synthetic_teleport}) comparing three different hypotheses: (i) uniform (solid, blue lines), (ii) structural (dashed, red lines) and (iii) popularity (dotted, purple lines). 
In each figure, the x-axis depicts the strength (weighting factor $k$) one assigns to a given hypothesis as defined in Equation~\ref{eq:chips} ($k=0$ refers to a flat prior) while the y-axis shows the corresponding evidence (marginal likelihood) value. For simplicity, we can compare hypotheses with each other by comparing the evidence values (larger values mean higher plausibility) for the same values of $k$  as all Bayes factors are decisive.
The results illustrate what we know from theory as for each dataset the hypothesis that captures the mechanisms according to which the data has been produced best, is declared as the most plausible one. 
}
\label{fig:synthetic}
\vspace{-10pt}
\end{figure*}

\section{Experiments}
\label{sec:experiments}
To demonstrate \approach and its general applicability, we perform experiments with both synthetic as well as empirical datasets (as introduced in Section~\ref{sec:data}).

\subsection{Experiments with Synthetic Data}
Our first experiments focus on
applying \approach to three synthetic trail datasets, generated by the following mechanisms:  teleportation, a random walk and a popularity random walk (see Section~\ref{sec:data}). We look at these three datasets and compare three corresponding hypotheses (uniform, structural, popularity) that capture the generative mechanisms of each dataset. 
As we know from theory, \approach ranks the hypothesis that best captures the underlying mechanisms as the most plausible one. 
%
Next, we introduce the hypotheses in greater detail, before we discuss the experimental results.


\smallskip
\noindent
\textbf{Hypotheses.}
We now describe how we express the three hypotheses as matrices $Q$. Note that we only have to specify the hypothesis matrix $Q$ (see Section~\ref{subsec:prior}) while the concrete pseudo count distribution for generating proper priors is handled by our approach.

\emph{\textbf{Uniform hypothesis.} }
This hypothesis has the intuition that trails have been purely generated by random teleportation and all transitions are equally likely. Thus, we equally believe in each transition and set each element of $Q$ to an equal value (here $1$). 

\emph{\textbf{Structural hypothesis.} }
This hypothesis captures our belief that the trails have been generated by (only) following the underlying topological link structure. Hence, we believe that agents would always choose a random link leading from one node to another while traversing the network. We express this by setting $q_{i,j}$ of $Q$ to $1$ if a directed link between state $s_i$ and state $s_j$ exists in the topological network.

\emph{\textbf{Popularity hypothesis.}}
This hypothesis also believes that the trails have been generated by following the links of the underlying link structure, but we have stronger beliefs in choosing large in-degree nodes compared to low in-degree nodes. Hence, we set $q_{i,j}$ to $\text{deg}^-(s_j)$ if a directed link between state $s_i$ and state $s_j$ exists in the topological network.


\smallskip
\noindent
\textbf{Results.}
Figure~\ref{fig:synthetic} depicts the results for each hypothesis and dataset at interest.
The x-axis denotes the weighting factor $k$ for the number of pseudo counts assigned (cf. Equation~\ref{eq:chips}). The y-axis denotes the corresponding Bayesian evidence (marginal likelihood). For $k=0$ the evidence is the same for all hypotheses as in that case the flat prior is uniformly distributed over the simplex and no additional informative aspect is considered.
 The larger $k$ gets, the more pseudo counts are assigned to the prior according to the given hypothesis and hence, the stronger our belief in specific transitions of a given hypothesis.
 We can compare hypotheses with each other by comparing the y-values (evidence) for the same x-values.
 According to Kass and Raftery's interpretation table of log-Bayes factors \cite{kass1995bayes}, we find that \emph{all differences are decisive} which is why we refrain from presenting explicit Bayes factors. Hence,  the larger the evidence for a given hypothesis is, the more plausible it is in comparison to the other hypotheses at interest. 
%
%
Across all three synthetic datasets, we can observe what we know from theory: the hypothesis that captures the underlying known mechanisms of the synthetic trails best is found to be the most plausible one. 
In the following we discuss the results of each dataset in detail:

\emph{\textbf{Structural random walk.}}
In Figure~\ref{subfig:synthetic_uniform} 
we can see that the structural hypothesis is ranked as the most plausible one as it exhibits the highest evidences for $k>0$. This result is as expected from theory as the trails are also produced according to a structural random walk only considering the underlying topological link network as expressed by the structural hypothesis. The reason why the popularity hypothesis is more plausible than the uniform hypothesis is because the former also incorporates structural information while the latter does not.

\emph{\textbf{Popularity random walk.}}
We show the results for our popularity random walk generated trails in Figure~\ref{subfig:synthetic_popularity}. In this case, the popularity hypothesis incorporating the in-degree (popularity) of potential structural target nodes is identified as the most plausible one as it captures the mechanisms according to which the trails have been generated.

\emph{\textbf{Random teleportation.}}
Finally, in Figure~\ref{subfig:synthetic_teleport} we demonstrate the results for our trails generated via random teleportation. As expected, the uniform hypothesis is the most plausible one which accounts to our prior belief that all target nodes are equally likely to come next given a current node.
Contrary, the structural and popularity hypotheses which both incorporate structural knowledge are less plausible hypotheses. 


\begin{figure*}[t!]
\vspace{-10pt}
\centering
\subfigure[Wikigame]{\label{subfig:wikigame}\includegraphics[width=0.33\textwidth]{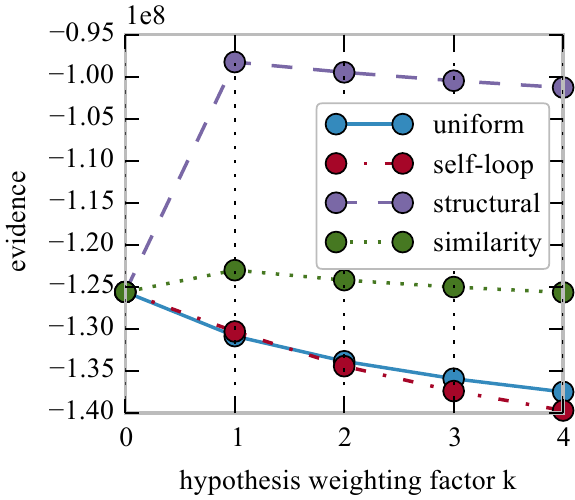}}
\subfigure[Yelp]{\label{subfig:yelp}\includegraphics[width=0.33\textwidth]{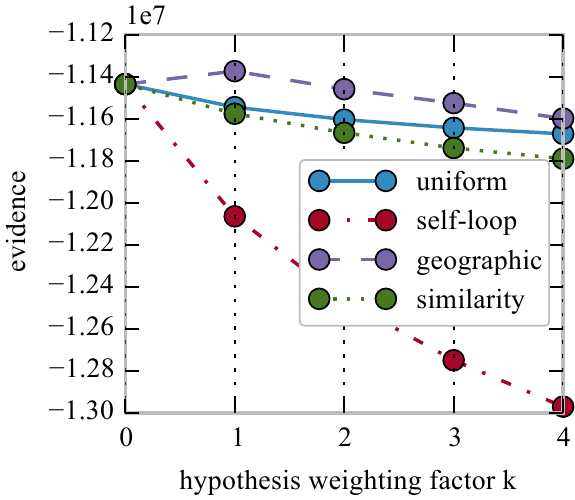}}
\subfigure[Last.fm]{\label{subfig:lastfm}\includegraphics[width=0.33\textwidth]{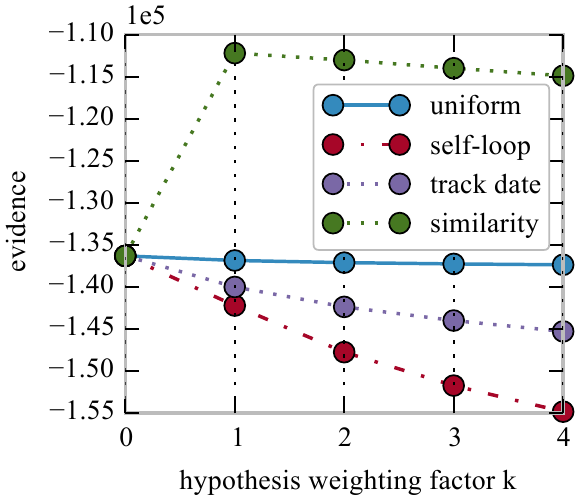}}
\vspace{-10pt}
\caption{\emph{Experiments with empirical data.} 
This figure depicts the results obtained from applying \approach to three different empirical human trail datasets (Wikigame \subref{subfig:wikigame}, Yelp \subref{subfig:yelp} and Last.fm \subref{subfig:lastfm}) for comparing a set of hypotheses.
The x-axis depicts the strength (weighting factor $k$) one assigns to a given hypothesis as defined in Equation~\ref{eq:chips} ($k=0$ refers to a flat prior) while the y-axis shows the corresponding evidence (marginal likelihood) value. For simplicity, we can compare hypotheses with each other by comparing the evidence values (larger values mean higher plausibility) for the same values of $k$ as all Bayes factors are decisive. 
Several domain-specific hypotheses are declared as the most plausible ones for our three datasets: the structural hypothesis for the Wikigame trails \subref{subfig:wikigame}, the geographic hypothesis for the Yelp trails \subref{subfig:yelp} and the artist similarity hypothesis for our Last.fm trails \subref{subfig:lastfm}.
%
%
}
\label{fig:empirical}
\vspace{-10pt}
\end{figure*}

\subsection{Experiments with Empirical Data}
\label{subsec:empirical}

Our second kind of experiments focus on demonstrating the general applicability of the 
\approach approach by applying it to three real-world, empirical human trail datasets (Wikigame, Yelp and Last.fm) as introduced in Section~\ref{sec:data}. We compare \emph{universal} as well as \emph{domain-specific} hypotheses for each dataset which we describe next, before we discuss the experimental results. 


\smallskip
\noindent
\textbf{Hypotheses.}
We now describe the universal and domain-specific hypotheses studied and how we express them. These are just exemplary hypotheses for illustrative purposes, researchers are \emph{completely free to formulate other / their own hypotheses} accordingly. 


\emph{\textbf{Uniform hypothesis.}} We use the universal uniform hypothesis in a similar fashion as for our experiments with synthetic data in order to express our prior belief that each state is equally likely given a current state. Hence, we assign $1$ to each element of the hypothesis matrix $Q$. We can see this hypothesis as a baseline for other hypotheses; if they are not more plausible than the uniform hypothesis, we cannot expect them to be good explanations about the behavior that is producing the underlying human trails.

\emph{\textbf{Self-loop hypothesis.}}
With the universal self-loop hypothesis we express our prior belief that humans never switch to another element in a trail. For example, for a navigational scenario this would mean that if a user currently is on a specific Wikipedia page, she would always just refresh the current one and never switch to another one. 
We set the diagonal to $1$ in the corresponding hypothesis matrix $Q$ and leave all other elements zero.

\emph{\textbf{Similarity hypothesis.}}
This hypothesis expresses our belief that humans consecutively target nodes in trails that are in some way (e.g., semantically) related to each other.
 We now aim at modeling this hypothesis for all three datasets. However, due to their given nature, the similarity hypothesis differs for each dataset which is why we describe the domain-specific similarity hypotheses next: 


\emph{Wikigame similarity hypothesis.} 
This hypothesis believes that humans prefer to consecutively access websites that are semantically related which has been observed and hypothesized in a series of previous works (e.g., \cite{singer2013computing,west,west3}). 
Using the set of Wikipedia pages that users navigate over, we use the textual information of each site provided by the corresponding Wikipedia dump (see Section~\ref{sec:data}) for calculating the semantic relatedness \cite{rubenstein1965contextual} between sites. We utilize a vector space model \cite{rubenstein1965contextual} for representing the documents (states) as vectors of identifiers using \emph{tf-idf} \cite{salton1988term} for weighing the terms of the vectors. 
We apply an automatic stop word removal process where we ignore all terms that are present in more than 80\% of the documents in our corpus. Also, we use sub-linear term frequency scaling as a term that occurs ten times more frequently than another is not necessarily ten times more important \cite{manning2008introduction}. Additionally, we perform a sparse random projection for reducing dimensionality while still guaranteeing Euclidian distance with some error \cite{li2006very,achlioptas2001database}. The final number of parameters is determined by the \emph{Johnson-Lindenstrauss lemma} \cite{dasgupta2003elementary} that states that given our $360,417$ number of samples (distinct Wikipedia pages), we only need $10,942$ features while preserving the results up to a tolerance of 10\%---which is also the tolerance level we use for dimensionality reduction. By doing so, we can reduce the number of tf-idf features from $2,285,489$ to the specified $10,942$. Finally, we calculate 
similarity between all pairs of pages (states) using \emph{cosine similarity} between the described vector representations which define $q_{i,j}$ of our matrix $Q$.
Each $q_{i,j}$ can now exhibit a final value between $0$ and $1$ where $1$ means complete similarity and $0$ means no relatedness at all. To increase sparsity we only consider similarities that are equal or larger than $0.1$. Additionally, we set the elements of the diagonal of $Q$ to $1$.


\emph{Yelp similarity hypothesis.} 
With this hypothesis we express our belief that humans choose the next business they review based on its similarity to the current business according to their categories (e.g., subsequently reviewing restaurants but not a barber after a restaurant).
 On Yelp, businesses can get assigned a list of categories that represent them (e.g., restaurant)---we leverage these categories for calculating similarity between businesses. Again, we use a vector space model for representing businesses as vectors of binary identifiers (category assigned or not assigned). For calculating all-pair similarity scores between businesses we utilize Jaccard similarity ranging from $0$ to $1$ which determine $q_{i,j}$ of the prior hypothesis matrix $Q$. The diagonal is set to zero as we do not believe in humans consecutively reviewing the same business.

\emph{Last.fm similarity hypothesis.}
This hypothesis believes that humans consecutively listen to songs on Last.fm if they are produced by the same artist---e.g., only listening to songs by Eros Ramazzotti.
We set elements of the hypothesis matrix $Q$ between two tracks to $1$ only if they are from the same artist---the diagonal is set to zero.

\emph{\textbf{Wikigame structural hypothesis.}}
For Wikigame, we evaluate an additional domain-specific hypothesis that captures our prior belief that users navigate the Web (or in this case Wikipedia) primarily by using the underlying topological link structure. The corresponding hypothesis matrix $Q$ can hence be built by looking whether links between sites of our states space $S$ exist in the underlying topological link network $G$ with directed edges $E(G)$ (derived from the Wikipedia dump as stated in Section~\ref{subsec:empirical_data}). To be precise, the values of the elements $q_{i,j}$ of $Q$ are determined by the number of hyperlinks linking from page $s_i$ to page $s_j$; mostly, only one hyperlink links from one page to the other. Additionally, we set the diagonal of the matrix to $1$ as users might also subsequently navigate the same page by e.g., clicking the refresh button of the browser.

\emph{\textbf{Yelp geographic hypothesis.}}
On Yelp, we also consider the domain-specific hypothesis that the next business a user reviews is one that is geographically close to the current one---we have used this as an example throughout this article (e.g., see Figure~\ref{subfig:hypo_geo} or Figure~\ref{fig:prior_viz}). 
For doing so, we start by calculating the \emph{haversine distance} \cite{sinnott1984virtues} between the longitude and latitude values of all pairs of businesses. As the resulting value (in km) is smaller for geographic close businesses than for far businesses we normalize the values by dividing them by the maximum distance before subtracting them from $1$. This leads to final values that range from $0$ to $1$ where $1$ means geographically identical. We set the values of $Q$ according to the calculated values while leaving the diagonal zero.

\emph{\textbf{Last.fm date hypothesis.}}
Finally, we specify a hypothesis that believes that successive tracks listened to on Last.fm are close regarding their original publication date (e.g., someone prefers to only listen to 80s songs). We determine the date of a track by using the Musicbrainz API and looking for the earliest release date available. Next, we calculate the difference between dates of two songs in years---we only consider track pairs for which we can retrieve a date for both tracks through the API. Similar to the Yelp dataset, we then divide each date difference value by the maximum and subtract it from $1$ giving us scores between $0$ and $1$ where the latter means that two tracks are originally published in the same year. We set the transition values of $Q$ according to the calculated values and leave the diagonal zero.


\smallskip
\noindent
\textbf{Results.}
The results for all datasets are shown in Figure~\ref{fig:empirical}. Again, all Bayes factors are decisive and we can simply interpret hypotheses having larger y-values (evidence, marginal likelihood) as more plausible. Across all datasets, we can identify some domain-specific hypotheses that are more plausible compared to the universal uniform hypothesis which we can see as a baseline. Hence, 
these hypotheses seem to capture some mechanisms well that human behavior exhibits while producing the human trails studied. 
%
Additionally, we find that 
throughout all datasets the universal self-loop hypothesis is the least plausible one with a small exception for the Wikigame dataset. We discuss the results for each dataset next.


\emph{\textbf{Wikigame.}}
First and foremost, the \approach results in Figure~\ref{subfig:wikigame}
show largest evidence for the domain-specific structural hypothesis.
This indicates that users playing the Wikigame indeed seem to prefer to navigate Wikipedia by following links of the underlying topological link network. This is not too surprising as the Wikigame per definition only allows users to click on available hyperlinks for trailing through the Wikipedia space. 
Additionally, we can see that the domain-specific similarity hypothesis is more plausible than both the universal uniform as well as the self-loop hypotheses. This corroborates the theories and assumptions of previous work \cite{singer2013computing,west,west3} which observed  that humans tend to follow semantically related concepts successively. 

We also see that the universal uniform and the self-loop hypotheses are the least plausible hypotheses at interest. Interestingly, for $k=1$ the self-loop hypothesis exhibits larger evidence than the uniform prior which partly also demonstrates that self-loops are indeed an existing aspect of this dataset as also observed in previous work \cite{singer2014detecting}. However, with larger $k$ the evidence of the uniform hypothesis surpasses the self-loop hypothesis which may be explained by the fact that we weigh the informative part (i.e., only self-loops) too strongly while we ignore all other possible transitions.

\emph{\textbf{Yelp.}}
We depict the results for comparing the hypotheses at interest for our Yelp dataset (business reviews) in Figure~\ref{subfig:yelp}. A first observation is that our approach indicates the domain-specific geographic hypothesis as the most plausible one. Hence, humans indeed seem to prefer to successively review businesses that are geographically close to each other on Yelp as captured by our dataset. Contrary, the other domain-specific similarity hypothesis is less evident compared to the uniform hypothesis which can be seen as a baseline. Consequently, from this exemplary analysis, we cannot assume that humans prefer to consecutively review the same businesses based on similar categoric descriptors, at least not based on the similarity of categorical descriptors given on Yelp. Finally, the self-loop hypothesis is indicated as the least plausible one which indicates that humans at maximum very seldom review the same business twice in a row in our dataset.

\emph{\textbf{Last.fm.}}
Finally, the Last.fm results depicted in Figure~\ref{subfig:lastfm}
highlight that the similarity hypothesis, expressing our prior belief that users consecutively listen to songs that stem from the same artist, is the most plausible one. This is visible as the evidence values are larger for all $k>0$ compared to the other hypotheses of interest. 
Again, we observe that humans do not seem to prefer to listen to the same song over and over again (self-loop hypothesis) in our dataset. Also, for this example data, the track date hypothesis is less plausible 
compared to the universal uniform hypothesis.




\section{Discussion}
\label{sec:discussion}


The \approach approach represents an intuitive way of comparing hypotheses about human trails as we have demonstrated on synthetic as well as empirical data. However, there are some aspects---as partly exhibited throughout our experiments---that one should consider when applying \approach; we discuss them next.


\smallskip
\noindent
\textbf{Expressing hypotheses.}
%
%
When expressing hypotheses as matrices,
choices have to be made regarding several factors such as (i) which transitions to believe in (e.g., about setting the diagonal) (ii) how to calculate values for representing a hypothesis (e.g., haversine distance for geographic closeness), (iii) how to normalize or (iv) availability of information (e.g., API restrictions). 
While several ways of doing this are conceivable, our approach does not constraint the researchers' choice in this regard.
If in doubt, our advice is to express the uncertainty through another hypothesis (which reduces the problem) or through a set of other hypotheses (which focus on different representations).
 For example, in this article we first 
investigate the plausibility of a universal self-loop hypothesis compared to a uniform hypothesis before making a choice about the diagonal of other hypotheses. We find that for navigational trails (Wikigame), self-loops seem to play a role at least occasionally (cf. Figure~\ref{subfig:wikigame})  which is why we also set the diagonals in other hypotheses to larger values than zero, while for both Yelp (cf. Figure~\ref{subfig:yelp}) as well as Last.fm (cf. Figure~\ref{subfig:lastfm}) we cannot observe such behavior. Another choice to make is whether one wants to express hypotheses in a symmetric or asymmetric way---e.g., it might be useful to believe that transitioning from state A to state B is more relevant than from B to A. Following our advice, we would express a symmetric and asymmetric version of the hypothesis and compare them.

\smallskip
\noindent
\textbf{Behavior of hypothesis weighting factor $k$.}
Throughout our experimental results (see Figure~\ref{fig:synthetic} and Figure~\ref{fig:empirical}), we frequently observe that the evidence is falling with larger $k$.
As pointed out in this article, the evidence is a weighted likelihood average and is largest if both the prior as well as the likelihood concentrate on the same parameter value regions. The larger we choose $k$, the larger we set the hyperparameters of the Dirichlet distributions and the more they get concentrated. Thus, only a few specific parameter configurations (single draw from the Dirichlet distribution) receive higher prior probabilities while many others receive low ones. As we cannot expect our hypotheses to concentrate on the exact same areas as the likelihoods as we did for our empirically aligned toy example in Figure~\ref{subfig:toy_empirical}, we sometimes see falling evidences with larger $k$ as we reduce the scope of the prior. 
Even though we always want to compare hypotheses with each other for the same values of $k$, comparing marginal likelihoods for different weighting factors $k$ might give further insights. For example, the flat prior for $k=0$ might be seen as a further baseline to compare to. Also, the behavior of the evidence for single hypotheses with varying $k$ can give insights into the strength we can assign to a given hypothesis.


\smallskip
\noindent
\textbf{Memoryless Markov chain property.}
Currently, \approach is memoryless, meaning that the next state only depends on the current one. Previous work has been contradictory in their statements about memory effects of human trails on the Web (see e.g., \cite{chierichetti,singer2014detecting}). While first order models have mostly been shown to be appropriate, it may be useful to extend \approach to also support memory effects in the future. This would mean that it would allow us to not only analyze hypotheses about how the current state influences the next one, but also how past ones (potentially) exert influence.

\smallskip
\noindent
\textbf{Future work.} 
In this work, we have suggested first ways of expressing hypotheses as well as eliciting Dirichlet priors. Future work should further refine these methods as well as study the sensitivity and interpretability of expressed hypotheses and elicited priors with respect to \approach; e.g., by focusing on how to normalize hypothesis matrices (potentially row-based) and refining methods for eliciting priors (distribution of chips).

Furthermore, while we have showcased a certain variety of datasets and hypotheses that can be analyzed with \approach, we would like to encourage researchers to see these examples only as a stepping stone for more detailed experiments to be conducted. Additionally, in future work, multiple extensions and / or experimental variations are conceivable.
For example, it could be useful to look at personalization or user group effects in data. Currently, the examples only demonstrate collective behavior, but one may assume that different groups of users produce human trails differently. One could segment the dataset according to some heuristic criteria and then analyze the same hypotheses on both sub-datasets. If one hypothesis is more plausible in one dataset than the other, one can assume differences in user behavior in different sub populations. One might also believe that human behavior changes over time \cite{yang2014finding}. This suggests to apply \approach to study the temporal evolution of hypotheses (and evidences for them). Furthermore, one can also think about combining hypotheses with each other to form new ones. For example, in Figure~\ref{subfig:wikigame} we show that both the structural as well as the similarity hypotheses are very plausible to explain navigational behavior on Wikipedia. One could use a combination of both by weighing structural transitions according to their similarity.

\section{Related Work}
\label{sec:relwork}


Studies of human trails in information systems have been fueled by the advent of the WWW \cite{berners2000weaving}. A fundamental way of interacting with the Web is navigating between websites. Such navigational trails have been extensively investigated in the past. An example of early work is by Catledge and Pitkow \cite{catledge} who studied navigational regularities and strategies for augmenting the design and usability of WWW pages. Subsequent studies, e.g., the work by Huberman et al. \cite{huberman} or Chi et al.~\cite{chi2001}, emphasize existing regularities and rationalities upon which humans base their navigational choices. These examples demonstrate the importance of better understanding sequential user behavior producing human trails on the Web. Apart from modeling \cite{singer2014detecting, pirolli,chi2001} and the detection of regularities and patterns \cite{huberman,walk2014cikm,walk2014discovering}, researchers have also been interested in studying strategies humans follow when producing human trails on the Web. We highlight some exemplary findings next.

A prominent theory is the \emph{information foraging theory} by Pirolli and Card \cite{pirolli99} which states that human behavior in an information environment on the Web 
is guided by \emph{information scent} which is based on the cost and value of
information with respect to the goal of the user \cite{chi2001}. Another behavioral pattern is shown in \cite{brumby} and \cite{pierce} where the authors observed that semantics affect how users search visual interfaces on websites; the importance of semantics between subsequent concepts is also emphasized in \cite{west,west3,singer2013computing,chalmers1998order}. Amongst many others, further studies of human trails on the Web have focused on the detection of progression stages \cite{yang2014finding}, trail prediction \cite{laxman2008stream}, the study of the value of search trail following for users \cite{white2010assessing,bilenko2008mining}, partisan sharing \cite{an} or the capture of trends in human trails \cite{matsubara2012fast}.

While we have highlighted just an excerpt of related work, all these studies reveal interesting behavioral aspects that could be translatable into hypotheses about transitions over states. What is difficult, is to compare them within a coherent research approach. In this work, we have tackled this problem. Fundamentally, \approach is based on a Markov chain model which is prominently leveraged for modeling human trails on the Web. Google's PageRank, for example, is based on a first order Markov chain model \cite{brin} and a large array of further studies have highlighted the benefits of Markov chain models for modeling human trails on the Web (e.g., 
\cite{singer2014detecting, walk2014discovering, deshpande, lempel, pirolli}). Given these advantages, and our interest in hypotheses about memoryless transitions, the Markov chain model represents a sensible choice for our approach. For deriving the parameters of models, we have utilized Bayesian inference \cite{singer2014detecting,Strelioff}.

The main idea of our approach is to incorporate hypotheses as informative Dirichlet priors into the Bayesian Markov chain inference and compare them with Bayes factors. Bayes factors are known to be highly sensitive on the prior. This property of Bayes factors has been seen as a limitation in the past---as originally pointed out by Kass and Raftery~\cite{kass1995bayes}.
However, as emphasized by Wolf Vanpaemel \cite{vanpaemel2010prior}, if "models are quantitatively instantiated theories, the prior can be used to capture theory and should therefore be considered as an integral part of the model". In such a case, the sensitivity of Bayes factors on the prior \emph{can be seen as a feature}---i.e., instrumental for gaining new insights into the plausibility of theories (or in our case hypotheses about human trails). Thus, marginal likelihoods and Bayes factors can be leveraged as an appropriate measure for evaluating hypotheses about human trails. The process of expressing theories as informative prior distributions over parameters has been discussed in follow-up work by Wolf Vanepaemel in \cite{vanpaemel2011constructing} and in \cite{vanpaemel2012using} where the author has tackled this task by using hierarchical methods.
In this work, we have presented an adaptation of the so-called (trial) roulette method, which was first proposed in \cite{gore1987biostatistics} and further discussed in \cite{oakley2010eliciting,Davidson2013}, for this task. 
With our adaption, we understand the grid as a hypothesis matrix where elements correspond to beliefs about transitions for a given hypothesis. Also, in our case, chips correspond to pseudo counts of Dirichlet priors which \approach automatically sets according to expressed hypotheses of researchers.

\section{Conclusion}
\label{sec:conclusion}

Understanding human trails on the Web and how
they are produced has been an open and complex challenge for our community for years. In this work, we have addressed a sub-problem of this larger challenge by presenting \approach, an approach that enables scientists to compare hypotheses about human trails on the Web. \approach utilizes Markov chain models with Bayesian inference. The main idea is to incorporate hypotheses as Dirichlet priors into the inference process and leverage the sensitivity of Bayes factors for comparing hypotheses. 
Our approach allows researchers to intuitively express hypotheses as beliefs about transitions between states which are then used for eliciting priors.

We have experimentally illustrated the general mechanics of \approach by comparing hypotheses about synthetic trails that were generated according to controlled mechanisms. As derived from theory, \approach ranks those hypotheses as the most plausible ones, that best capture the mechanisms of the underlying trails.
Additionally, we have studied empirical data to further show the general applicability of \approach. We have looked at human trails from three different domains: human navigational trails over Wikipedia articles (Wikigame), successive reviews of businesses (Yelp) as well as trails capturing songs that users consecutively listen to (Last.fm). Although the experiments presented in this work mainly serve to illustrate how one can apply the \approach approach, we hope that they also motivate and encourage researchers to conduct further, more in-depth studies of human trails on the Web in the future. 

While we have developed \approach for comparing hypotheses about hypertext trails, the approach is not limited to Web data. It can be applied to any form of trails over states at interest in a straight-forward manner;  e.g., it could also be used to study human trails as recorded by GPS data. Insights gained by such studies can give a clearer picture of the underlying dynamics of human behavior that shape the production of human trails.




\smallskip
\noindent
\textbf{Acknowledgements.}
This work was partially funded by the DFG German Science Fund research project "PoSTs II" and by the FWF Austrian Science Fund research project "Navigability of Decentralized Information Networks". Furthermore, we want to thank Alex Clemesha for giving us access to the Wikigame data as well as Daniel Lamprecht and Paul Laufer for valuable input.



\clearpage
\small
\bibliographystyle{abbrv}
\bibliography{biblio}

\end{document}